# A Novel Local Binary Pattern Based Blind Feature Image Steganography

Soumendu Chakraborty[1]* and Anand Singh Jalal[2]

[1] Department of Information Technology, Indian Institute of Information Technology, Lucknow, U.P., INDIA
[2] Department of Computer Engineering and Application, GLA University, Mathura, U.P., INDIA
E-mail addresses: soum.uit@gmail.com*, anandsinghjalal@gmail.com
*Corresponding author. mobile number:+919897043787

Abstract: Steganography methods in general terms tend to embed more and more secret bits in the cover images. Most of these methods are designed to embed secret information in such a way that the change in the visual quality of the resulting stego image is not detectable. There exists some methods which preserve the global structure of the cover after embedding. However, the embedding capacity of these methods is very less. In this paper a novel feature based blind image steganography technique is proposed, which preserves the LBP (Local binary pattern) feature of the cover with comparable embedding rates. Local binary pattern is a well known image descriptor used for image representation. The proposed scheme computes the local binary pattern to hide the bits of the secret image in such a way that the local relationship that exists in the cover are preserved in the resulting stego image. The performance of the proposed steganography method has been tested on several images of different types to show the robustness. State of the art LSB based steganography methods are compared with the proposed method to show the effectiveness of feature based image steganography.

Keywords— LBP, LSB, Steganography, Feature based steganography.

## 1. Introduction

Image steganography is the art of hiding secret message in a cover image in such a way that the presence of secret data is not detectable [1]. Primarily there are two types of image steganography schmes; Reversible [18-26] and Irreversible. Many LSB (Least Significant Bit) based image staganography [2-3] methods have been proposed to hide the secret message in LSB of the grayscales of the cover image. These methods are prone to easy detection, if exposed to strong feature based image steganalysis.

LSB matching (LSBM) [2] makes some modifications in embedding process and improves the visual quality of the stego image. Instead of embedding the secret bit directly to the LSB of the cover, LSBM adds 1 to the LSB of the cover if it matches with the present secret bit and subtracts 1 from LSB otherwise. LSBM ties to preserve the structure of the cover image by maintaining almost equal number of odd and even intensities. However, LSBM cannot sustain strong feature based stego analysis. This method is almost similar to LSB steganography. LSB matching revisited (LSBMR) [3][4] is an improved version LSBM, which reduces the embedding rate to preserve the local structure. LSBMR embeds a single bit into a pixel pair of the cover image. The relationship of the pixel pair is used to embed the secret bit, which preserves the local relationship of the pixel pair. As only one pixel of the pair is modified to embed two secret bits the modification rate reduces to 0.375 bit per pixel (bpp) [3]. The proposed method is better than LSBMR in a sense that it preserves the local structure of the larger neighborhood. There are some edge adaptive methods proposed in literature such as hide behind corner (HBC) [5]. Image steganography using edges to embed the secret bit in selected edge areas can significantly improve the embedding capacity [6][7]. Edges are computed as the difference between pixels. These edge pixels are used to embed the data. A secret key has been used to rotate the original cover image and secret bits are embedded into the edge areas of the modified cover using LSBMR by Luo et al. [6]. The method proposed in [7] identifies edge and non-edge pixels of the cover image after removing 5 least significant bits of each pixels intensities. More pixels are embedded in the non-edge pixels to achieve more embedding capacity. Another LSB based non blind image steganography method has been proposed in [8], where adaptively secret data is embedded into the edge area of a grayscale image. Most recently a blind LSB base steganogrpahy technique has been proposed by Jiang et al. [9]. Quantum images are computed to hide the secret data in the LSB of the image [9]. Pixel value differencing (PVD) based method proposed by Swain [10] adaptively embeds the secret message into the selected area of an image. LSB and PVD based method proposed in [11] classify the cover images into high and low texture regions and embeds secret bits using PVD and LSB method and then the embedding is improved by recursive application of PVD shift and MPE embedding. There are PVD based methods where secret bits are adaptively embedded with respect to the different pixel value differences [16-17]. Most of these methods though complex fails when tested with strong feature based steganalysis methods. The reason being significant change in local structure of the stego image with increasing embedding rates. In this paper LBP (Local Binary Pattern) [12] based steganography is proposed which preserves the local structure of the cover image in the resulting stego image. The LBP based steganography over Haar wavelet has been proposed in [30]. In this method LBP operator has been applied over wavelets of the cover image. The embedding capacity of this method is very less. The method proposed in [30] achieves only 62dB Peak Signal to Noise Ratio [PSNR] [28] for embedding capacity of 1920 bits, where as the proposed method achieves 58.64dB PSNR for 822423 bits. Most recently deep learning based networks (encoders) are being used to hide the payload image [31-32]. These networks prepare the secret image according to the cover then the cover and the secret images are encoded using the encoder. These methods are highly complex and dependent on exhaustive training.

The feature based descriptor is used to hide the data in [33] where scale invariant feature transform (SIFT) is used to hide the information. Integer Wavelet Transform (IWT), block division, and Center Symmetric Local Binary Pattern (CSLBP) have been used to embed data in [34]. The method proposed in [34] achieves 67.40dB PSNR for a bit rate of only 0.0156, whereas our method achieves 58.64dB PSNR for a staggering bit rate of 3.37.

## 2. Proposed method

The proposed method is a unique blind image steganography method that preserves the local statistical structure and effectively hides the payload using the binary statistics of the cover image. To embed the payload it is first converted into 8 bits and Local Binary Pattern (LBP) is extracted from the local neighborhood of the cover image. These two binary values are X-ORed to compute the X-ORed value, which is shuffled pair wise to generate final binary string. This binary string is embedded into the cover and each converted pixel of the cover is synchronized using (8) to preserve the local structure of the cover in the stego image. The proposed extraction is the exact reverse process of embedding as shown in Fig.1(b).

### 2.1 Embedding Process

Proposed steganography method explores the local relationship among the pixels of the cover image and encodes the same using LBP. LBP is a binary pattern computed by comparing the center pixel of a 3×3 mask with 8 neighboring pixels. This 8 bit pattern represents the visual characteristics of the image and changes with non uniform environmental changes. Flow diagram of embedding process is shown in Fig.1(a). The cover image is denoted by $C$ of size $N \times N$, which is divided into non-overlapping $3 \times 3$ blocks. $C_{i,j}$ denotes the reference pixel of each block, whose local neighborhood is encoded using LBP. Pixel $C_{i,j}$ is encoded using (1), where $f(.)$ is a binary function.

$$LBP_{i,j} = \{f(C_{i,j}, C_{i+0,j+1}), f(C_{i,j}, C_{i-1,j+1}), f(C_{i,j}, C_{i-1,j+0}), f(C_{i,j}, C_{i-1,j-1}), f(C_{i,j}, C_{i+0,j-1}), f(C_{i,j}, C_{i+1,j-1}), f(C_{i,j}, C_{i+1,j+0}), f(C_{i,j}, C_{i+1,j+1})\} \quad (1)$$

$f(.)$ is defined as

$$f(C_{i,j}, C_{i+a,j+b}) = \begin{cases} 1, & if\ C_{i,j} \geq C_{i+a,j+b} \\ 0, & else \end{cases} \quad (2)$$

Where $a, b \in \{-1,0,1\}$

The encoded LBP is used to modify the pixels $P_{k,l}$ of the secret image (payload) of size $M \times M$ (where $N = 3 \times M$) using XOR operation as shown in (3).

$$X_{k,l} = LBP_{i,j} \oplus P_{k,l}|_{i=2k, j=2l} \quad (3)$$

The bits of the modified pixel $X_{k,l}$ of the payload denoted as $x_{k,l}^n$ (please note that n denotes the bit position and bit position 0 represents least significant bit) are embedded into the local neighborhood of $C_{i,j}$ to compute the stego image. The center pixel (reference pixel) $C_{i,j}$ is kept unchanged to preserve the local relationship among the neighborhood pixels as shown in (4). It is absolutely necessary to conserve the local relationships, as the proposed method is a blind steganography method

$$S_{i,j} = C_{i,j} \quad (4)$$

Eight bits of $X_{k,l}$ consist of 4 pairs, each pair with two adjacent bits. In every pair the two bits are swapped. The resultant pixel $Y_{k,l}$ is given as

$$Y_{k,l} = \{y^0_{k,l} = x^1_{k,l}, y^1_{k,l} = x^0_{k,l}, y^2_{k,l} = x^3_{k,l}, y^3_{k,l} = x^2_{k,l} \ldots y^6_{k,l} = x^7_{k,l}, y^7_{k,l} = x^6_{k,l}\} \qquad (5)$$

The most significant bit $y^7_{k,l}$ of $Y_{k,l}$ is embedded into the right side neighbor $C_{i+0,j+1}$ of $C_{i,j}$ at the least significant bit position using (6).

$$S_{i+0,j+1} = \sum_{a=1}^{7} 2^a \times c^a_{i+0,j+1} + y^7_{k,l} \qquad (6)$$

The second most significant bit $y^6_{k,l}$ of $Y_{k,l}$ is embedded into the upper right side neighbor $C_{i-1,j+1}$ of $C_{i,j}$ at the least significant bit position using (7).

$$S_{i-1,j+1} = \sum_{a=1}^{7} 2^a \times c^a_{i-1,j+1} + y^6_{k,l} \qquad (7)$$

Similarly, remaining bits of $Y_{k,l}$ are embedded into the remaining 6 neighbors of $C_{i,j}$. To preserve the local neighborhood relationship stego image pixels must be synchronized. A synchronization function $sync(.)$ is defined in (8) to preserve the local neighborhood relationship.

$$sync(C_{i,j}, C_{i+a,j+b}, S_{i+a,j+b}) = \begin{cases} S_{i+a,j+b} - (2^\mu), & if(C_{i,j} \geq C_{i+a,j+b}) \text{ and } (C_{i,j} < S_{i+a,j+b}) \\ S_{i+a,j+b} + (2^\mu), & if(C_{i,j} < C_{i+a,j+b}) \text{ and } (C_{i,j} \geq S_{i+a,j+b}) \\ S_{i+a,j+b}, & \text{otherwise} \end{cases} \qquad (8)$$

Where $\mu = 1,2,3 \ldots$ are the number bits flipped/inserted within one pixel of the cover image. For example, if only one bit is inserted per pixel then, $2^\mu = 2$.

## 2.2 Extraction Process

Secret image is extracted directly from the stego image using LBP. Fig.1(b), shows the extraction process of the proposed scheme. As the local neighborhood relationship is preserved during embedding process, the LBP can be extracted from stego image $S$ using (1). The bits of the shuffled pixel $Y_{k,l}$ of the payload are extracted from the local neighborhood of each reference pixel $S_{i,j}$ of the stego image. Modified intensity $X_{k,l}$ are recovered from shuffled pixel $Y_{k,l}$ using (9). Finally, the actual pixels of the secret image (Payload) are extracted using (10).

$$X_{k,l} = \{x^0_{k,l} = y^1_{k,l}, x^1_{k,l} = y^0_{k,l}, x^2_{k,l} = y^3_{k,l}, x^3_{k,l} = y^2_{k,l} \ldots x^6_{k,l} = y^7_{k,l}, x^7_{k,l} = y^6_{k,l}\} \qquad (9)$$

$$P_{k,l} = LBP_{i,j} \oplus X_{k,l}|_{i=2k, j=2l} \qquad (10)$$

The example shown in Fig. 2 encodes a sample intensity and embeds the resulting shuffled bits into the cover. It is evident from the embedding example shown in Fig.2 that the LBP feature is preserved in the local neighborhood of the stego image. Fig.3 shows the extraction process where the pixel of the payload is extracted correctly from the local neighborhood of the stego image.

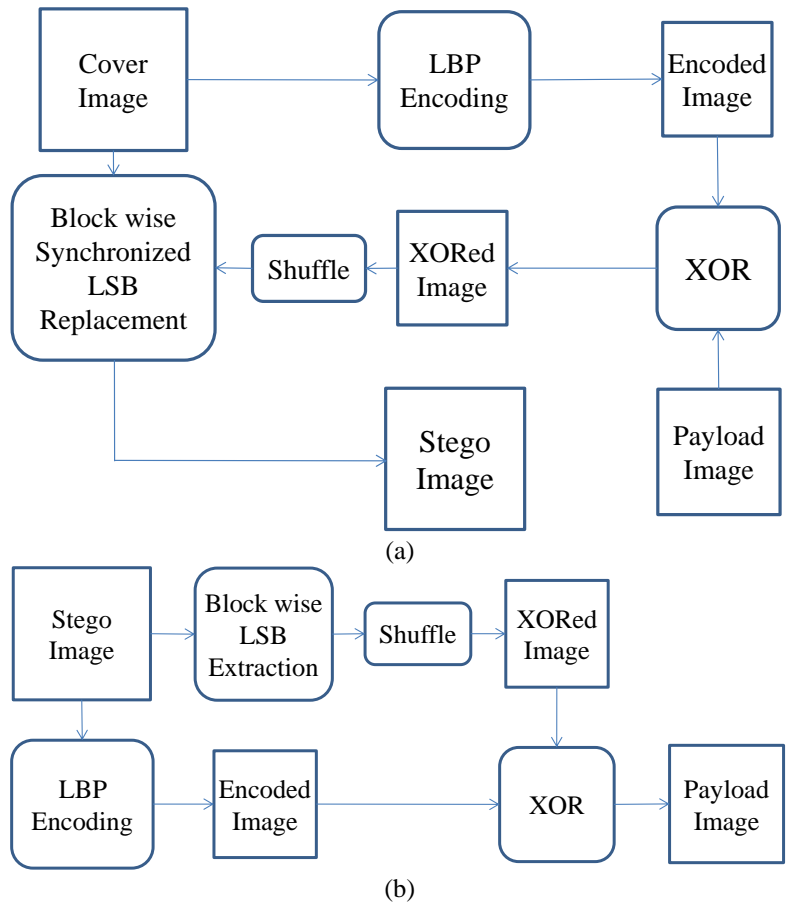

Figure 1: Proposed method; (a) Embedding process, (b) Extraction process

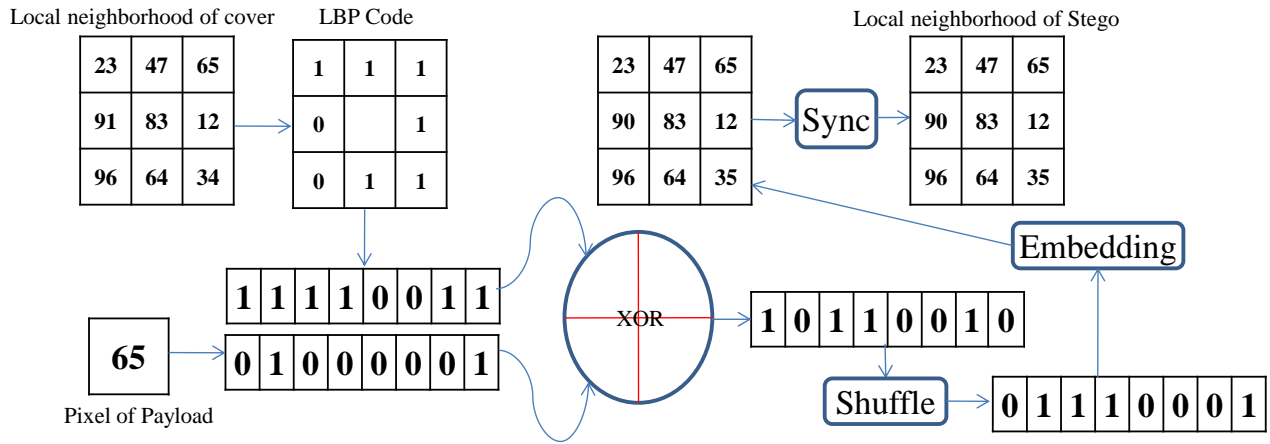

Figure 2: An example showing the embedding process of one pixel of the payload.

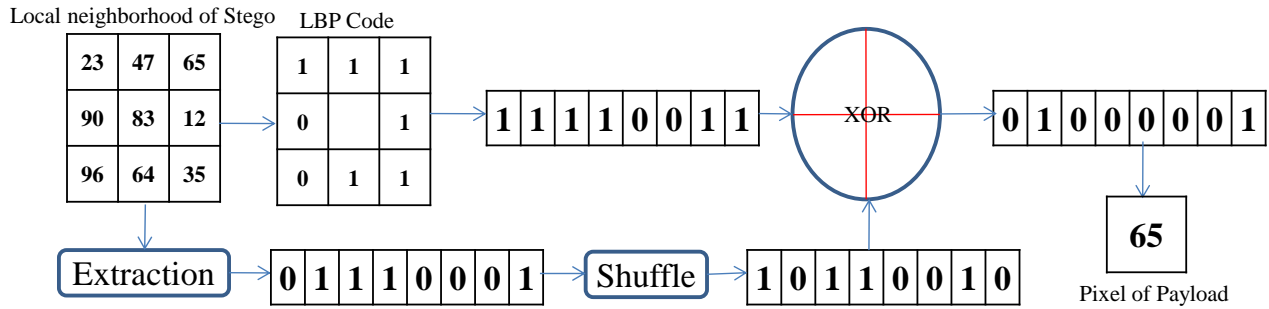

Figure 3: An example showing the extraction process of one pixel of the payload.

*2.3 Complexity Analysis*

The proposed method computes LBP code with 8 comparisons to embed a single pixel of the payload. It takes 8 bit wise X-OR operations to compute modified pixel of the payload and 4 circular shifts of length 2 bit to compute the shuffled pixel. It requires at most 8 additions/subtractions to replace the LSB of the cover to generate the intermediate stego pixel intensities. To synchronize the stego pixels, the proposed method requires at most 8 additions/subtractions. Hence, the proposed method requires at most 28 fundamental operations to embed a single pixel of the payload. To embed a payload of size $M \times N$ the number of fundamental operations is at most $28 \times M \times N$. So, the computational complexity of the embedding process is $O(M \times N)$. Similarly, the extraction process requires 20 fundamental operations to extract a single pixel of the payload. Therefore, the computational complexity of the proposed extraction process is also $O(M \times N)$. It is evident from the complexity analysis of the proposed method that it computes the stego image with comparable complexity.

*2.4 Major Contribution*

The steganography methods embed the secret bits into the cover image, which drastically affects the local structure of the image. Local relationships are dependent on the visual quality and characteristics of the image such as illumination, shape of the objects, background etc. These characteristics get transformed by insertion of the secret bits. Strong feature analyses are used to identify these changes, which can easily recognize the presence of secret information in the cover image. So far in our study, we have not come across such steganography methods, which preserve these local structures representing the characteristics of the cover image. The proposed method computes one such binary structure and embeds the secret information in the cover image in such a way that this local binary structure is preserved. The need and importance of this method has been validated through extensive analysis in section 3.

## 3.   Performance analysis

To show the robustness of the proposed method three different types of analysis has been done. The visual quality of the stego image has been analyzed by computing the histogram. Quantitative measure such as PSNR has also been used to show that the proposed method maintains the visual quality of the cover image after embedding. Most frequently used statistical features are computed and analyzed in sections 3.3, 3.4, and 3.5 to show functional impact of the feature based steganography.

*3.1 Qualitative Analysis*

Most fundamental statistical feature of any image is the histogram. The original histogram of a cover and the histogram of the corresponding stego image with different embedding rates have been shown in Fig. 4. Histogram of the stego image even with 50% embedding is similar to the histogram of the cover image. This shows that the proposed method effectively preserves the statistical features of the cover image.

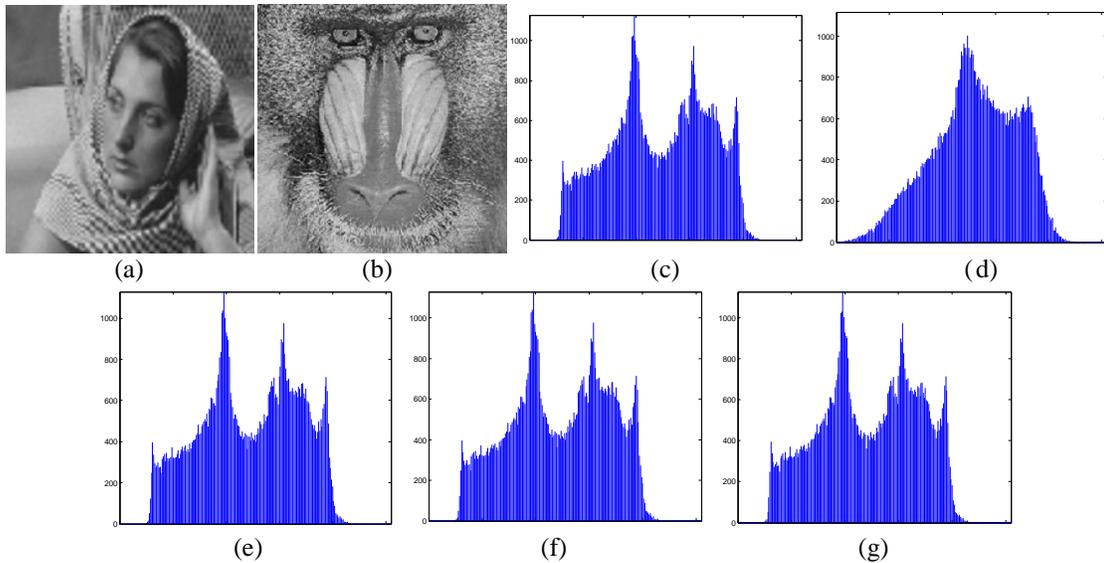

Figure 4: (a) Cover image, (b) Payload, (c) Histogram of cover, (d) Histogram of payload, (e) Histogram of stego image with 30% embedding, (f) Histogram of stego image with 40% embedding, (g) Histogram of stego image with 50% embedding.

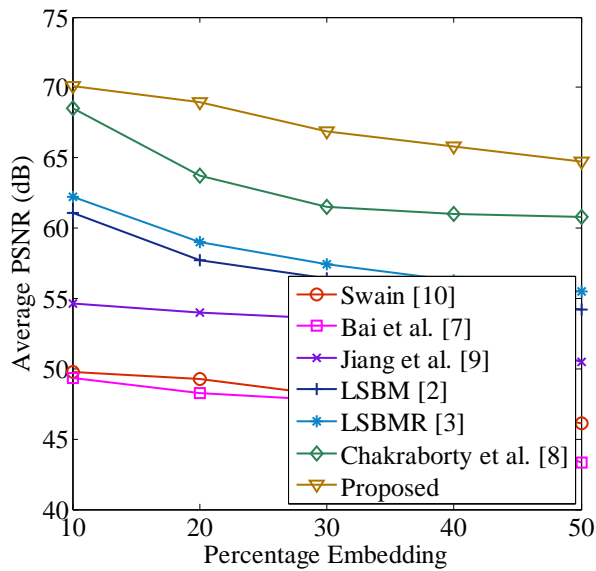

Figure 5: Average PSNR of different methods for different embedding rates.

## 3.2 Quantitative Analysis

To show the visual quality of the stego images, 206 different types of images from USC-SIPI database [13][27] has been used. Each of the images taken as payload (secret image) in turn is embedded, with different embedding rates, into the rest of the images to compute the stego images using state of the art methods as well as the proposed method. Mean of the PSNR of these stego images corresponding to a single payload is then computed and finally average of these mean values corresponding to all payloads is computed to get the average PSNR [28]. Average PSNR of different methods are shown in Fig.5. The proposed method shows at least 6dB improvement in average PSNR over state of the art methods.

The proposed method has also been tested on most challenging images listed in Fig.6. It has been compared with the state of the art PVD based LSB replacement methods [16-18] with respect to Embedding Capacity (EC) [16-17], PSNR, Bit Rate (BR) [16-17], and Quality index (Q) [16-17][29]. The proposed method has been compared with seven-way

PVD of [16], and Type-1 technique of [17]. The comparative results are shown in Table 1. The proposed method achieves significantly better embedding capacity with low bit rate and very high PSNR and good quality index.

### 3.3 Statistical R-S Analysis

RS analysis [8] has been used to test the effect of increasing embedding rates. Regular and singular groups are identified with respect to a mask $m$. Mask $m$ is a set of -1, 0 and 1which captures the flipping of pixels of the cover image. General idea of the RS analysis is to detect the change in regular and singular groups with increasing embedding rates. To avoid detection of the presence of secret message in stego images difference between regular groups $R_m$, $R_{-m}$ and singular groups $S_m$, $S_{-m}$ should be restricted to minimum. Fig.7 shows the RS diagram for the proposed method for two images interchangeably taken as cover and payload. As shown in Fig.7 the difference between regular singular groups remains constant with increasing embedding rates. Hence the probability of detection of the secret message is minimal.

### 3.4 Statistical Analysis with Pixel Difference Histogram

Pixel Difference Histogram (PDH) is one of the most frequently used statistical tools to identify the stego images. In this section PDH of the original images and corresponding stego images with 50% embedding are computed and shown in Fig.8. In PDH of all the stego image step effect is not visually detectable. The PDH shown for Barbara Stego image and Elaine Stego image are almost coincided with the PDH of the original image. Even though the points of occurrences of Lena Stego and Baboon Stego do not coincide with the corresponding PDH of the original image, they are collinear. Fig.8. shows that the proposed method is robust against PDH analysis even with 50% embedding.

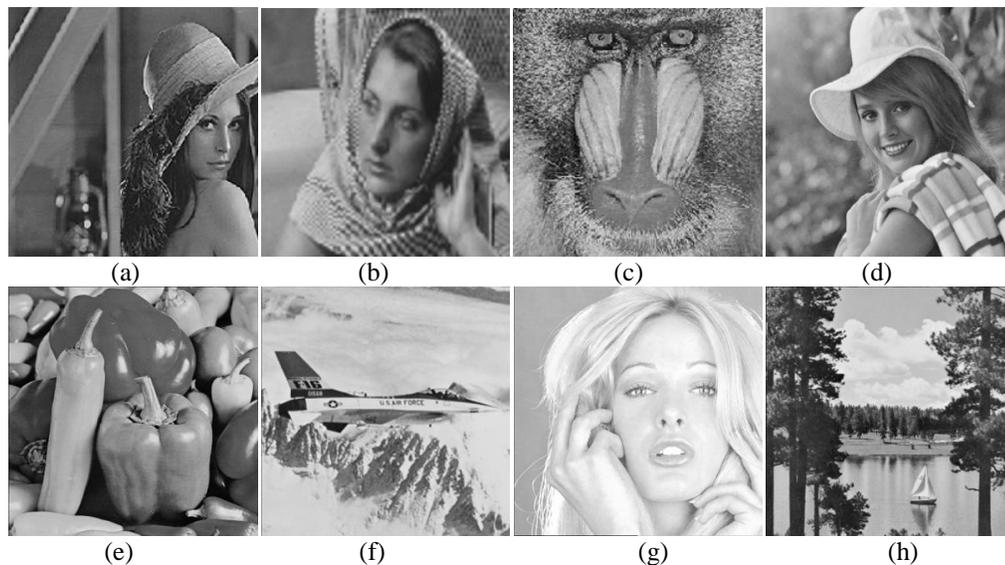

(a)　　　　　　　(b)　　　　　　　(c)　　　　　　　(d)

(e)　　　　　　　(f)　　　　　　　(g)　　　　　　　(h)

Figure 6: Selected original images used in some of the statistical and quantitative analysis (a) Lena, (b) Barbara, (c) Baboon, (d) Elaine, (e) Peppers, (f) Jet, (g) Tiffany, (h) Boat.

Table 1: Comparative analysis with respect to Embedding Capacity (EC), Bit Rate (BR), Quality index (Q), and PSNR

| Grayscale Image | Method [16] | | | | Method [17] | | | | Proposed ($\mu = 4$) | | | |
|---|---|---|---|---|---|---|---|---|---|---|---|---|
| | EC | BR | Q | PSNR (dB) | EC | BR | Q | PSNR (dB) | EC | BR | Q | PSNR (dB) |
| Lina | 632220 | 2.41 | 0.9993 | 41.76 | 791749 | 3.02 | 0.9995 | 41.25 | 832106 | 3.41 | 0.9998 | 56.82 |
| Baboon | 742268 | 2.83 | 0.9957 | 33.79 | 814453 | 3.10 | 0.9982 | 34.49 | 885488 | 3.99 | 0.9973 | 53.57 |
| Tiffany | 466918 | 1.78 | 0.9986 | 41.23 | 790798 | 3.01 | 0.9980 | 40.25 | 862333 | 3.52 | 0.9995 | 58.25 |
| Peppers | 592690 | 2.26 | 0.9993 | 40.42 | 790952 | 3.01 | 0.9991 | 37.91 | 822423 | 3.37 | 0.9995 | 58.64 |
| Jet | 635418 | 2.42 | 0.9989 | 42.09 | 791349 | 3.01 | 0.9986 | 40.89 | 847238 | 3.46 | 0.9993 | 56.23 |
| Boat | 657362 | 2.50 | 0.9988 | 37.89 | 797341 | 3.04 | 0.9993 | 40.05 | 846754 | 3.47 | 0.9987 | 57.95 |

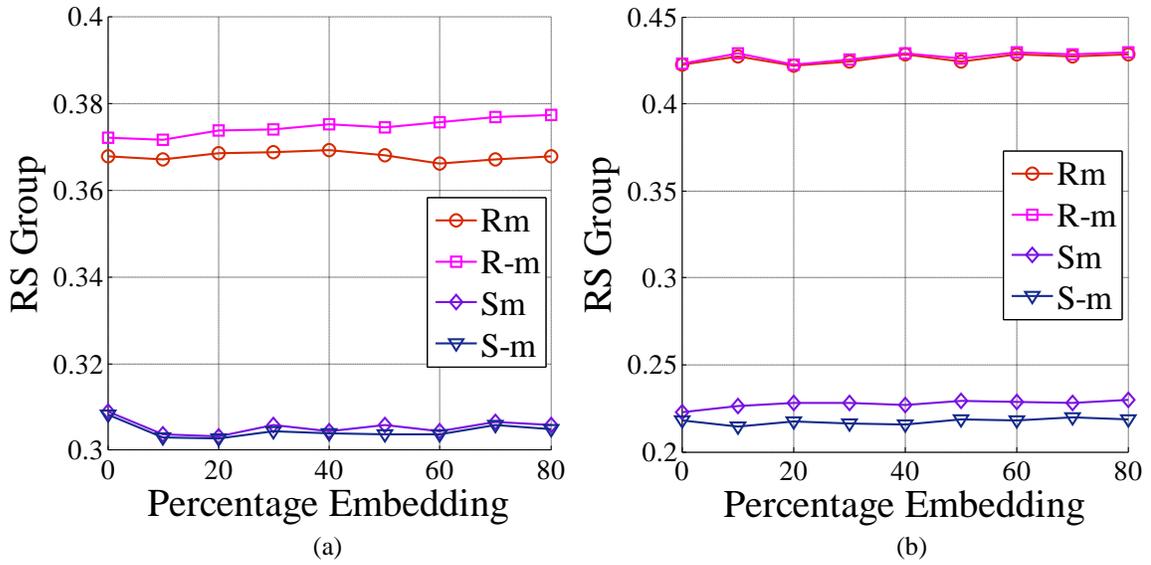

Figure 7: RS diagram with (a) baboon as cover and Lena as payload, (b) Lena as cover and baboon as payload.

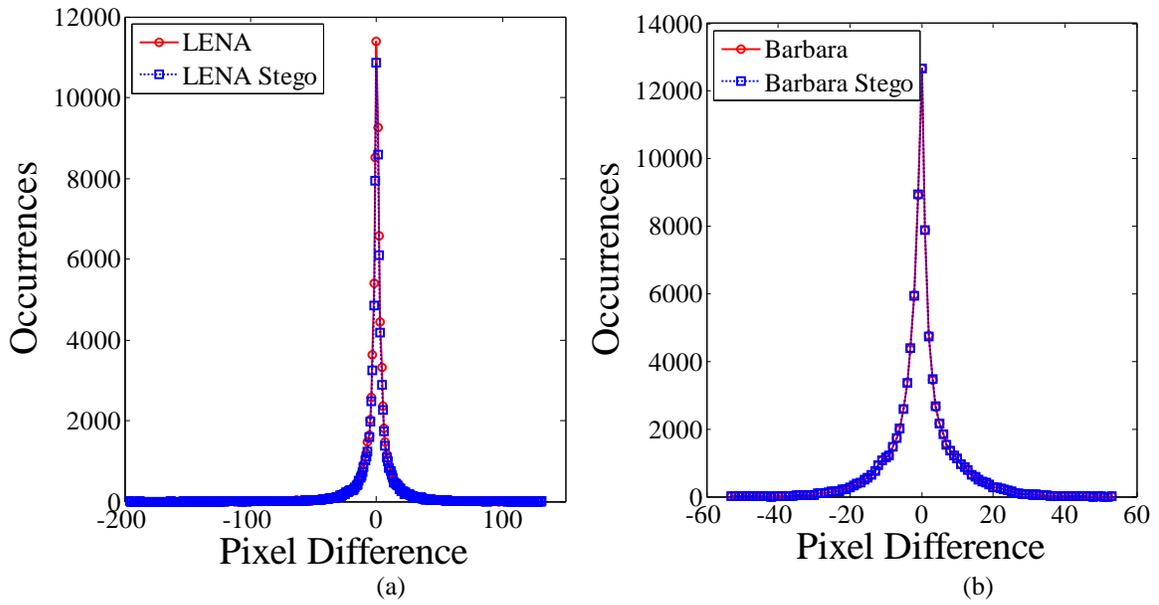

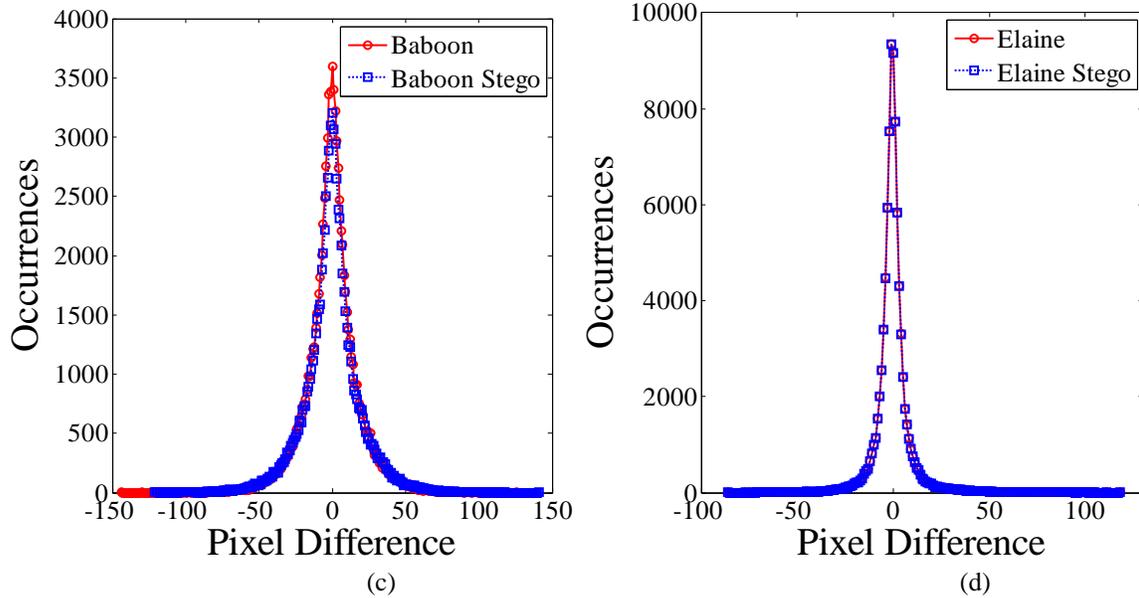

Figure 8: Pixel Difference Histogram with 50% embedding (a) Lena, (b) Barbara, (c) Baboon, (d) Elaine.

### *3.5 Higher Order Feature Based Analysis*

The proposed method is also tested and compared using a strong Ensemble classifier [14] over USC-SIPI database[13][27]. Stego images are computed for each of the images in the database as elaborated in the previous paragraph. The Gabor features [15] are computed for each of the cover and the corresponding stego image for different embedding rates. A 2-dimensional Gabor filter [15] shown in (11) has been used to compute the Gabor response with two scales and two directions (0°, 90°). A Gabor filter is a Gaussian response over sinusoid with frequency *f* and standard deviations $\sigma_s$ and $\sigma_t$ [16].

$$\emptyset(s,t) = \frac{1}{2\pi\sigma_s\sigma_t} e^{[-(1/2)(s^2/\sigma_s^2 + t^2/\sigma_t^2) + 2\pi i f s]} \quad (11)$$

Gabor feature of a stego image for a particular embedding rate is taken as test set and the remaining Gabor features of the stego and corresponding cover images are taken as training set to train the Ensemble classifier. If the classifier correctly classify a stego image with a particular embedding rate then it is taken as 1 otherwise 0. Total number of correct classifications for a particular embedding rate is the classification accuracy of a particular steganography scheme. It is desirable that the classification accuracy should be less for a robust steganography scheme. As shown in Fig. 9, the classification accuracy for the proposed scheme is at least 2% less than the nearest counterpart chakraborty et al. [8].

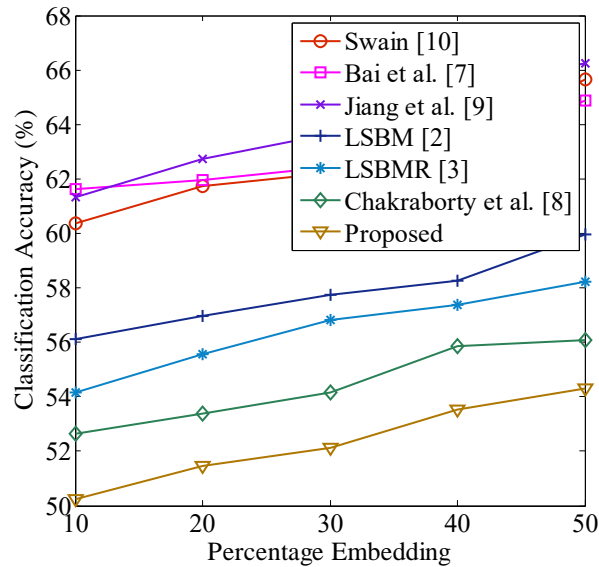

Figure 9: Average classification accuracy with ensemble classifier and Gabor feature.

## *4. Conclusion*

The proposed scheme is a unique feature based robust steganography technique which preserves the local structure of the cover into the resulting stego image. As shown through various performance evaluation experiments, the proposed scheme out performs most recent state of the art steganographic schemes. State of the art steganography fractures the local structure of the cover and makes it possible to detect the presence of hidden data through strong statistical feature based steganalysis. The proposed method effectively preserves the local structure and proves to be robust against feature based steganalysis as shown by the experiments. The proposed work can be extended by introducing payload specific hand crafted descriptors to embed the data. The hand crafted descriptors must be analyzed against the secret data first, then the suitable descriptor should be chosen.